\documentclass[psfig]{aa}
\usepackage{graphicx}
\begin{document}

\title{Interactions, mergers and  
the fundamental  mass relations of galaxies}

\author{Patricia B. Tissera
          \inst{1},\inst{2},
	Anal\'{\i}a V. Smith Castelli
          \inst{1},\inst{3},
          \and
           Cecilia Scannapieco
          \inst{1},\inst{2}
          }

\offprints{P. B. Tissera}

\institute{Consejo Nacional de Investigaciones Cient\'{\i}ficas y T\'ecnicas, Argentina
              \email{patricia@iafe.uba.ar}
         \and
            Instituto de Astronom\'{\i}a y F\'{\i}sica del Espacio, Argentina
         \and
              Facultad de Ciencias Astron\'omicas y Geof\'{\i}sicas, La Plata, Argentina}

\abstract{
We present a study  of the effects of mergers and interactions on the mass distribution of galactic systems in
 hierarchical clustering scenarios using 
the disc-bulge structural parameters and their dynamical properties to quantify them. We focus
on the analysis of the Fundamental Mass Plane  relation, finding that
secular evolution phases contribute significantly to the determination of a plane with a slope in agreement to
that of the  observed luminosity relation. In these simulations, secular phases are responsible for 
the formation of  
compact stellar bulges with the correct structural parameter
combination. We also test that the relations among these parameters
agree with observations. The Kormendy mass relation is also reproduced after
secular evolution phases.
From our findings, we predict that 
the departure of systems from the $z=0$ Fundamental Mass Plane, involving a change in the slope, 
 could indicate a lack of
secular evolution in their formation histories.  
Taking into account  these results, the hierarchical growth of the structure
predicts a bulge formation scenario for typical field spiral
galaxies where secular evolution during
dissipate mergers plays a fundamental role. Conversely,   subsequent mergers
can help to enlarge the bulges but do not seem to  strongly modify
their       fundamental mass relations.
Systems get to the local mass relations at different stages of evolution (i.e. different redshifts)
so that their formation histories introduce a natural scatter in the relations.
We also found that the parameters of the  Tully-Fisher Mass relation for the disc components are correlated with those of the
 Fundamental mass one for the bulge components at least during mergers events, so that as the systems increase
its circular velocity, the bulges get more concentrated.
Our results suggest that  the formation mechanisms  of the 
 bulge and  disc components,  satisfying their corresponding
 fundamental mass relations, might be coupled and that 
 secular evolution could be the possible connecting process.


}

 \authorrunning{Tissera et al}

   \titlerunning{Effects of interactions on the fundamental relations. }

   \maketitle

\section{Introduction}

The origin of the fundamental relations of galaxies such as the Tully-Fisher (TF) 
for spiral
galaxies and the Fundamental Plane (FP) for ellipticals and early-type bulges, is 
still under discussion.
Although,  assuming the virialization
of the systems, the main features of both relations can be explained within a hierarchical scenario
, mergers and interactions
redistribute mass and angular momentum in galaxies and, as a consequence, they can affect the relation between
their structural and dynamical parameters.
 
The simplest scenarios currently accepted for the formation of bulges in disc 
galaxies are two: the first one establishes that the bulge 
and the disc are formed independently, the bulge being formed previously to the 
disc (Andredakis, Peletier \& Balcells 1995); the 
second scenario assumes that the disc forms first and, the bulge emerges from it as a 
consequence of gas inflows during a period of secular evolution (Courteau, de Jong 
\& Broeils 1996). 
Within hierarchical clustering scenarios, bulges have been modeled
through the merger of smaller systems (e.g. White \& Rees 1978; 
Cole et al. 2000). However, 
 it is still a matter of  debate 
which are the main physical  mechanisms involved in the formation of bulges and at what extent they are coupled
to the formation of the disc component.

Observationally, several works have considered  to  this problem. 
By applying a bulge-disc decomposition with a S\'ersic profile to bulges, Andredakis et al. (1995) found that 
the S\'ersic index $n$ varies in a continuous and smooth way from values 
$n \approx 1$ for late-type bulges, to values $n\approx 6$ for early-type 
bulges. As a consequence, the authors argued that all bulges formed by a common 
mechanism (see also Bender, Burstein \& Faber 1992; Khoroshahi et al. 2000ab; M\"ollenhoff \& Heidt 
2001). However, Courteau et al. (1996) suggested that such a smooth sequence could be the result of a scenario
in which early-type bulges emerged from a minor merger, and late-type bulges form via secular
evolution, both processes operating to some degree of efficiency for all Hubble types.

The formation of disc systems is well-explained by assuming the conservation of the specific angular momentum 
of the baryonic component as it collapses onto the dark matter potential well
of the system (Fall \& Efstathiou 1980; White \& 
Rees 1978). This simple scheme has served as a basis to build more complex models for galaxy formation (e.g.
Lacey et al. 1993; Mo, Mao \& White 1998), which have been successful in reproducing several observational
results, but which have the shortcoming of not being able to   self-consistently describe 
the effects of mergers and interactions, ubiquitous in the current cosmological paradigm.

In fact, if the structure forms in a hierarchical scenario, mergers
play an important role in the formation of galaxies and, as a consequence,
in the determination of their properties and fundamental relations.
Numerical simulations remain a useful tool to study galaxy formation in a cosmological
context.
Regarding the  FP relation of spheroids, Capelato, de Carvalho \& Carlberg (1995) and Evstigneeva et al. (2003)
have investigated the effects of mergers by  using pure  N-body numerical simulations.
Recently, Kobayashi (2005) studied the formation of elliptical galaxies in cosmological simulations,
finding that the different merger histories of the systems introduced an important scatter in their
FP relation. 

The formation of disc systems in hierarchical scenarios have been studied by several authors.
In particular, Scannapieco \& Tissera (2003, hereafter ST03) 
studied the effects of mergers on
the structural properties of disc-like systems
 by using smooth particle 
hydrodynamical (SPH) numerical simulations.
These authors focused on the analysis of the mass distributions of disc-like objects and
how they are modified during mergers. In this work, a  comparative study
of the effects of secular evolution triggered by tidal fields and of the actual
collision of the baryonic clumps was carried out. 
ST03 found that galactic objects formed in 
hierarchical clustering scenarios reproduce the angular momentum and structural
parameter distributions of spiral galaxies,  if a stellar compact 
bulge is allowed to
form and early gas depletion is avoided.
The formation of a compact  stellar bulge provides stability to the cold gaseous discs (Athanassoula \& Sellwood 1986; Binney \& Tremaine 1987;
Barnes \& Hernquist 1991, 1992; Martinet 1995 and references therein;
Mihos \& Hernquist 1994, 1996; Mo, Mao \& White 1998),
preventing the triggering of  non-axisymmetric instabilities 
by interactions and mergers (Christodoulou, Shlosman \& Tohline 1995; van den Bosch 1998, 2000).
 The latter can  cause important angular momentum  losses
followed by strong gas inflow (Navarro \& Benz 1991; Navarro, Frenk \& White 1995ab; Evrard, Summers \& Davis 1994; Vedel, Hellsten \&
Sommer-Larsen 1994; Navarro \& Steinmetz 1997; Sommer-Larsen, Gelato \& Vedel 1999; 
Weil, Eke \& Efstatiou 1998).
By adopting an artificial low efficiency for the star formation activity, ST03 (see also Dom\'{\i}nguez-Tenreiro, Tissera \& S\'aiz 1998
and S\'aiz et al.  2001) were able to avoid the formation of too- concentrated, pure gaseous discs 
(e.g. Navarro \& Benz 1991; Navarro, Frenk \& White 1995ab; Evrard, Summers \& Davis 1994) or a stellar
spheroid-dominated system (e.g. Thacker \& Couchman 2000). 
 Supernova energy feedback should provide a self-consistent regulation of the star formation activity, allowing
the formation of extended discs. However, a self-consistent SN feedback model within SPH has not been yet presented 
(see Scannapieco et al. 2006 for a new promising approach).

In this paper, we will continue the analysis of ST03,
investigating other structural parameters and, particularly,
the   
FP relation, with the aim of analyzing how mergers change the mass distribution 
and the fundamental relations. 
Contrary to pre-prepared mergers, we use fully
self-consistent, cosmological
hydrodynamical simulations, where the
distribution of
merger parameters
and the physical characteristics of the interacting systems
% the masses
%of the virial halos and baryonic clumps involved in the merger event, and the
%spin, internal structure, and relative orientation of the baryonic clumps
%that are about to merge, among others, 
arise naturally at {\it each
epoch\/} as a consequence of working within a  cosmological scenario.
No ad hoc set of orbital parameters or dynamical characteristics for the galactic systems need
to be assumed in this work.
%of the initial spectrum of the density fluctuation
%field, its normalization, and the cosmological models and its parameters.
The effects of mergers and
interactions on star formation can be studied in a consistent
scenario, although at the expenses
of losing numerical resolution.
%We also analyzed three high resolution mergers to test possible numerical effects.

This paper is organized as follows. In Sect. 2 we give a summary of the simulations and
method used by ST03 to determine the structural parameters. In Sect. 3
we discuss the results and compare them  with observations. Sect. 4 summarizes
the results.

\section{Numerical experiments}

The simulations used in this paper have been analyzed by Dom\'{\i}nguez-Tenreiro, Tissera \&
S\'aiz (1998), Tissera (2000) 
 S\'aiz et al. (2001), Tissera et al. (2002), and ST03.
These simulations are consistent with a standard Cold Dark Matter 
 ($\Omega =1$, $\Lambda =0$, $H_0 = 100 {\rm km s^{-1}\ Mpc^{-1}}\ h^{-1}$ with $h=0.5$) universe and
 represent a  5 $h^{-1}$ Mpc side box  using 
 $64^3$ particles.  The gravitational softening adopted is   
$1.5$ kpc $h^{-1}$, and the minimum hydrodynamical smoothing length
is  $0.75$ kpc $h^{-1}$. A baryonic density parameter
of $\Omega_{\rm b}=0.10$ is assumed. 
All baryonic and dark matter particles have the same 
mass, $M_{\rm p}=2.6\times 10^{8} {\rm M}_{\odot}$.
The three realizations of the power spectrum (cluster normalized) were run by using 
 the SPH code of Tissera, Lambas \& Abadi  (1997), which includes radiative 
cooling and star formation (hereafter, main runs).
In this paper, we also analyzed three typical galactic systems in a high resolution 
run performed with Gadget-2 (Sect. ~\ref{resolution})
 to assess possible numerical effects.

 The size of the simulated  volume is the result of a compromise between
the need to have a well-represented galaxy sample and
enough numerical resolution to study the astrophysical properties of the
simulated galaxies.  We are confident that since 
we  focus our analysis on small-scale processes such
as mergers and interactions, scale fluctuations of the order of
the box size  will have
no significant effect on such  local processes.
Moreover, since we are interested in the possible changes triggered by galaxy-galaxy interactions, the results
are also  weakly dependent on the detail cosmological parameters adopted (see also P\'erez et al. 2006).
In this work, we aim to understand the physical processes at work during mergers and
not to probe the cosmological model itself.
As we mentioned before, we focus this study on the effects that mergers have on
the mass distributions of the galactic objects,
 analysing them as individual events and
not in connection with their history of evolution or environment.

The star formation algorithm used in these models is   based
on the Schmidt law and transforms cold and dense gas particles into
stars if  they are denser than a certain critical value and
satisfy the Jeans instability criterion (Navarro \& White 1994).
Gas particles are checked to satisfy these conditions at all
time-steps of integration. Then, as the gas cools down and accretes onto
the potential wells of  dark matter halos, it is
gradually transformed into stars.
The star formation timescale of a given gas particle is assumed to be proportional
to its dynamical time by adopting a star formation efficiency parameter. This parameter
 is the only free one in these models and has been kept constant
in all analyzed simulations.
 
The star formation efficiency used in these simulations is the one adopted
by S\'aiz et al. (2001),
since it is adequate to reproduce disc-like structures with
observational counterparts at $z=0$. These authors used a low
star formation efficiency which
allows the formation of  stellar bulges that  assure the
axisymmetrical  character of the potential well, but
without exhausting the gas reservoir of these systems (see also 
Dom\'{\i}nguez-Tenreiro et al. 1998).
As a consequence, disc-like structures can be formed, although
they remain mainly gaseous.  If the star formation process 
 continued normally, the gaseous  discs would  be  transformed into
stars. These stars would  inherit
 the dynamical and astrophysical properties
of their progenitor gas clouds. Hence, 
we assume that structural properties of the gaseous discs
reflect those of the stellar discs that formed out of them. 
In this paper, we will determine and use these parameters to study the effects
of mergers on the mass distribution of the simulated galactic objects
and the fundamental relations they define.

\subsection{Merger events and bulge-disc decompositions}

We used the set of merger events studied by ST03.
These authors 
 identified galactic objects at their virial radius, analyzing
only those  with more than 4000 total  particles at $z=0$  within their
virial radius  to diminish numerical resolution problems.
 Within each galactic object, a main baryonic
clump is individualized, which will be, hereafter,  called the galaxy-like
object (GLO).

The selected GLOs have   very
well-resolved dark matter halos
 which provide adequate  potential wells
 for baryons to collapse in. This fact assures a reliable description of
the gas density profiles (see Steinmetz \& White 1997), which allows us
to follow the star formation history of the GLOs (see Tissera 2000).
With this strong restriction on the minimum number of particles, the
final GLO sample at $z=0$ is made up of 12 GLOs with virial velocities in
the range 140-180 km$\cdot$s$^{-1}$.
These systems spawned from $z=0$ to $z\approx 1.5$, as shown by ST03.

ST03 followed the evolution of the selected GLOs back in time,
constructing their merger trees and star formation rate (SFR)
 histories.
Then   mergers and starbursts (SBs) were identified. 
In this paper, we  worked with the same set of GLOs and merger events analyzed by ST03.
The latter are classified taking into account the SF activity taking place during the mergers.
 During a merger event,  the progenitor
object is chosen  as the more massive baryonic clump within this merger tree,
while the minor colliding baryonic
clump is referred to as the satellite.
If no gas inflows are induced during the
orbital decay phase (ODP), then   only one
SB is triggered when the two baryonic clumps collide. These events are cataloged as single SBs (SSBs). 
If the system is unstable and tidal torques are able to
drive instabilities inducing early gas inflows during the ODP, then  two SBs are detected, one
associated with the ODP and the second one with the actual collision. These events are classified as 
double SBs (DSBs).
The  total set of events compromises  18 mergers  with orbital and dynamical parameters settled by
the cosmological model. Among them, 11 are classified as DSBs and 7
as SSBs.

Merger events are characterized  by four redshifts of 
reference; $z_{\rm A}$: the beginning of the first bursts in DSBs 
and the ODP in SSBs, $z_{\rm B}$: the end of both the first bursts in
DSBs and the ODP in SSBs, $z_{\rm C}$: the beginning of the second SBs
in DSBs and the only bursts in SSBs (in this case $z_{\rm B}=z_{\rm C}$),
and $z_{\rm D}$: the end of the second bursts in DSBs and of the only one
in SSBs. % Fig.\ref{zABCD} (adapted 
%from ST03) shows two typical cases of SSBs and DSBs where these redshifts of reference have been included.
 It is important to stress
that the ODP phase is determined by
$z_{\rm A}$ and $z_{\rm C}$. These two redshifts determined  the  time elapsed since 
 two objects share the same dark 
matter halo until the fusion of their main baryonic cores. The $z_{\rm B}$ redshift is only  used in the case
of DSBs to measure  the end of the first SBs.

We used the structural parameters calculated by ST03,
who 
carried out a bulge-disc decomposition
of the integrated projected baryonic mass surface density of the GLOs at $z=0$ and
of the progenitors at each of the redshifts that define
a merger event ($z_{\rm A},z_{\rm B},z_{\rm C}$, and $z_{\rm D}$). 
The projections were carried out on the plane perpendicular to the total angular momentum of the systems.
ST03 followed the procedure explained in detail by S\'aiz et al. (2001). 
ST03 adopted   the S\'ersic law (S\'ersic 1968) for the bulge central mass concentration and an exponential for the disc.
As discussed in S\'aiz et al. (2001), it is more robust to perform the fitting of the integrated mass density
distributions since they are not binning-dependent and are less noisy than the density profiles.

Note that in numerical studies  it is  mass density distributions and not the luminosity density profiles
which are analyzed.
In this work we are interested in the the changes in the mass distribution
during mergers but for the sake of comparison with observations we need to assume mass-to-light ratios.
 We adopt a disc mass-to-light ratio $\Gamma =7$ regardless of redshift, disc scalelenghts and observational band and a ratio between the disc and the bulge
mass-to-light ratios of 20 to be consistent with the analysis of
ST03. The combination of these mass-to-light ratios provides a good
fit to the fundamental relations.
Note, however, that we are using the total baryonic masses to estimate luminosities (due to our low star formation
efficiency). In a more realistic model including Supernova feedback a fraction of this mass should be
heated up and even expelled from the systems. Hence, the adopted mass-to-light ratios should be considered only as
upper limits.

We carried out 75 disc-bulge decomposition (see S\'aiz et al. 2001 and ST03 for details
of the procedure and error discussion). The  baryonic  mass
 and the circular velocity ($V_{2.2}$) are  measured at $2.2 r_{\rm d}$. The analyzed systems can be
classified as  intermediate spirals ($100 \ {\rm km \ s^{-1}} <V_{2.2}< 180 \ {\rm km \ s^{-1}}$, 46 systems), late spirals 
($ V_{2.2}> 180 \ {\rm km \ s^{-1}}$ and $r_{\rm d} > 5.25 $ kpc, 6 systems), or compact spirals 
($ V_{2.2}> 180 \ {\rm km \ s^{-1}}$ and $r_{\rm d} < 5.25 $ kpc, 23 systems). Because of our minimum required number of particles, we do not
have any dwarf galaxies in this sample. 

\subsection{Observational data for comparison}

Since in these simulations we are analyzing the properties of the mass distributions, the 
 confrontation of  the results with observational data is not direct.
 While  observations provide information on luminosity density distributions,
our analysis will give us 
insight into the mass properties 
that are not affected by  dust or stellar  evolution.
 However, we would like to compare our findings with observations, and to 
 this end, we assume  mass-to-light ratios for the bulge and for the disc components  to match the fundamental relations at $z=0$. But the reader should baer in mind that the comparison with observations is made in broad terms.

We compiled observations from different authors who analyzed the
brightness surface density profiles (re-scaled to $H_0=50 {\rm ~km~s^{-1}~Mpc^{-1}}$) by using the S\'ersic law
to fit the bulge brightness distribution. 
 For the  structural parameters, we used the data from  
 Khosroshahi et al. 2000b (K00b) which include 26 early-to intermediate-type spirals in the K-band.
% and  Graham \& Guzman (2003, hereafter GG03)
%observed  18 dE galaxy candidates in the Coma Cluster in  F606W-band. 
These authors used the same software packages that we applied in this work to perform the bulge-disc fittings (i.e. MINUIT).
For the luminosity FP, we adopted the derived by 
Falc\'on-Barroso, Peletier \& Balcells (2002, hereafter FBPB02) 
from  19 S0-Sbc bulges in the K-band. Note that all these observations correspond to galaxies at z $\approx$ 0.
Unfortunately, we do not have a unique set of observed parameters to confront the structural parameters and
the fundamental plane.

\subsection{Numerical resolution}
\label{resolution}

Regarding numerical resolution, in the main runs, the baryonic components of galaxy-like systems are resolved with a
relatively low number of particles. In contrast, dark matter halos are
described with a much better resolution. An inappropriate low gas resolution
would result in a nonphysical gas heating that could halt the gas collapse.
In fact, some works suggest that it
is an inadequate resolution in the dark matter halo component that may produce
the larger  numerical artifacts (Steinmetz \& White 1997).
In fact, it appears that a well-resolved dark matter halo, no matter if
the number of gas particles is low (although larger than 500 gas particles), gives rise to a well-represented
gas density profile, this being  the most important point for
both the hydrodynamics and the tracking of the star formation history.
Dom\'{\i}nguez et al. (1998) have already proved that it is possible to reproduce 
populated and extended discs in high resolution simulations of standard CDM scenarios.

In this paper,  we  carried out a high resolution simulation by using
$2 \times 64^3$ total particles (hereafter HR)  with a chemical  Gadget-2 (Scannapieco
et al. 2005),  adopting
{\it the same cosmology and initial conditions of one of our main runs} (i.e., the parameters of the high resolution simulation
are the same listed in Sect. 2 for the main runs, except for the particle masses).
 Dark matter is resolved with the same number of particles (i.e. $64^3$) in the main 
and high resolution runs, but the  baryonic resolution is improved by one order of magnitude in the HR experiment (i.e. $64^3$).
 The star formation efficiency was set, as we
did in the main runs,so that stars formed only in the very high density regions.
 In this way, compact stellar cores are allowed to form without 
exhausting the gas reservoir at early stages of evolution (Dom\'{\i}nguez et al. 1998; ST03). The bulge/disc parameters obtained
for the high resolution systems  are consistent with the relations obtained from the lower resolution run.

It deserves to be stressed that the high resolution run is performed with a different
code, Gadget-2, which has been improved to conserve entropy when appropriate and includes chemical
evolution and metal-dependent cooling. Hence, the fact the galactic systems follow the trends
 obtained for those in our main runs  provides  strong support to our findings.

\section{Results}

In this section, we  assess the structural parameters of the simulated bulges and the discs and the fundamental
 relations
they determine during merger events and at  $z=0$. 
We have adopted a mass-to-light relation in order to confront the results  with observations. But,  we 
are mainly interested in the mass redistribution during merger events and how that affects possible correlations.
 Furthermore, the correct estimations of luminosities would
imply the treatment of other physical processes such as SN energy and metallicity feedback which are beyond the scope
of this work.

\subsection{Correlations among structural parameters}

\begin{figure}
\begin{center}
\vspace*{-0.2cm}\resizebox{8.5cm}{!}{\includegraphics{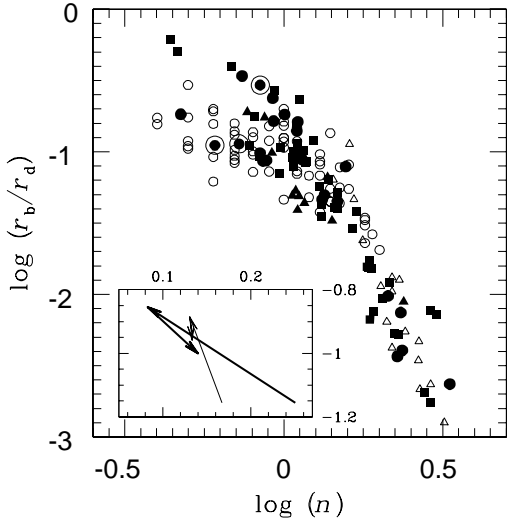}}\vspace*{-0.2cm}%
\end{center}
\caption{Ratio between bulge and disc scale lengths as a function
of the bulge shape parameter  for simulated objects at
$z=0$ (filled triangles), single SBs (filled circles), and double
ones (filled squares). Filled circles in open circles correspond to the parameters for
the high resolution systems.
Observational data
from Khosroshahi et al.  (2000b; open triangles) and
MacArthur, Courteau \& Holtzman  (2003; open circles) are also shown.
The small box shows the changes of the plotted parameters for single
(solid thin line)  and double (solid thick line) SBs during the
orbital decay phase (first arrow) and the fusion of the baryonic clumps
(second arrow).}
\label{fig-paperI}
\end{figure}

In this section we focus only on the analysis of the structural parameters of the bulges and the
relation with those of the discs, since Dom\'{\i}nguez et al. (1998), S\'aiz et al. (2001), Tissera et al. (2002), and ST03 
had exhaustively studied the properties of
 disc-like structures.
In these works, the simulated GLOs were found to have  bulge and disc  scale lengths
comparable to those of spiral galaxies and to show the 
expected relation with the shape parameter $n$, as 
discussed by ST03. 
In  Fig.~\ref{fig-paperI}, 
we show  the 
ratio between the bulge and the disc scale lengths ($r_{\rm b}$, $r_{\rm d}$) of the 
simulated objects as a function of 
the shape parameter (Fig. 9 in ST03). We have included the values for the GLOs analyzed
in the HR test, which  are found to be consistent with the relation determined by the main runs.

\begin{figure}
\begin{center}
\vspace*{-0.2cm}\resizebox{7.5cm}{!}{\includegraphics{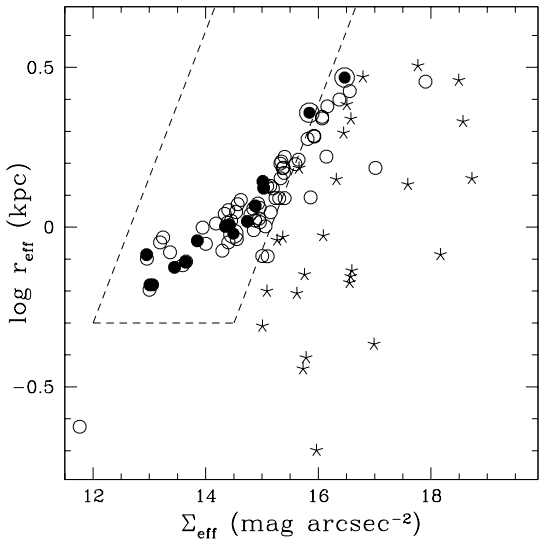}}\vspace*{-0.2cm}%
\end{center}
\caption{
$Log~n$ as a function of
 $\Sigma^{\rm o}_{\rm b}$ for simulated bulges
 at
$z \leq 0.03$ (filled circles) and $z > 0.03$ (open circles).
Filled circles in open circles correspond to the parameters for
the high resolution systems.
We have included the correlation
coefficients.
 The observational data from   Khosrhoshahi et al. (2000b, stars) are
also displayed for comparison.}
\label{nvsSigma0}
\end{figure}

We have investigated possible correlations between the shape
parameter $n$ and the structural parameters,
finding only a signal of anti-correlation with $\Sigma^{\rm o}_{\rm b}$,
as shown in 
 Fig.~\ref{nvsSigma0} (see also Table 1).
It is interesting to note that the simulated bulges yield the same
slope for this relation independently of the redshift. 
%although with a larger scatter for GLOs at $z >0$. 
In Fig.~\ref{nvsSigma0} we have also included the observations
from K00b which include  early and intermediate spirals.
We can appreciate a slight offset between simulations and observations which could be due to
the fact that we are using a fixed  mass-to-light ratio and total baryonic masses to estimate
surface brightness. 
The HR systems  are consistent with the trend determined by
GLOs in the main runs
(for the HR systems, we assume the same mass-to-light ratios used in the main runs).

\begin{figure}
\begin{center}
\vspace*{-0.2cm}\resizebox{7.5cm}{!}{\includegraphics{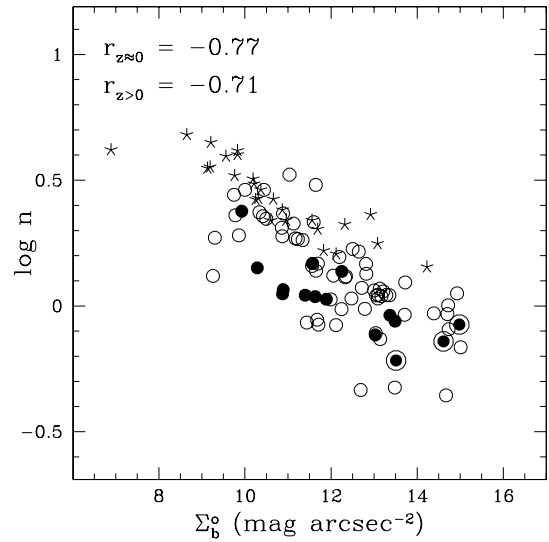}}\vspace*{-0.2cm}%
\end{center}
\caption{Bulge effective surface 
density ($\Sigma_{\rm eff}$) as a function of the effective radius 
($r_{\rm eff}$),
for simulated objects 
at $z \leq 0.03$ (filled circles) and $z > 0.03$ (open circles). 
 Filled circles in open circles  correspond to the parameters for
the high resolution systems at $z=0$.
Observational data
from Khosrhoshahi et al. (2000b, stars) are shown for comparison.
We also depicted the region  occupied by  ellipticals from 
 Khosrhoshahi et al. (2000a).
}
\label{grahamtotI}
\end{figure}

In   Fig.~\ref{grahamtotI}, we show  the  
bulge effective surface density ($\Sigma_{\rm eff}$) as a function of 
the effective radius ($r_{\rm eff}$)
for all simulated objects at all studied redshifts and for the HR systems. We used the
relation given by 
MacArthur et al. (2003) to derive $\Sigma_{\rm eff}$ and $r_{\rm eff}$.
% from 
%$\Sigma^{o}_{\rm b}$ and $n$.
To carry out a more thoughtful analysis, we performed linear
regressions through the data and the observations.
In Table 1  we show the slopes of the  fits and the corresponding 
  Pearson linear correlation coefficient, $r$.
As it can be appreciated, there are   clear correlation signals 
 between the simulated  structural parameters
of the bulges (log $r_{\rm eff}$  vs. $\Sigma_{\rm eff}$)
of galaxy-like objects  
at $z\approx 0$ and for their progenitors ($z > 0$).
 We found that the slope of the simulated relation at $z=0$ agrees with
the one derived from K00b within the observed dispersion.
The same is valid for the relation estimated for the simulated bulges
at $ z > 0$, although, in this case,  we note a change in the slope for small $r_{\rm eff}$. We will come back to this point in Sect. 3.2  
The fact that these simulated parameters for the bulges show a tighter relation that that of Fig.~\ref{nvsSigma0}
is produced by their definations,   which are coupled through the 
 a shape parmeters $n$. %\footnote{agregar definicion}.
In the case of $n$ and $\Sigma^{o}_{\rm b}$, they are both  estimated from the bulge-disc 
decomposition as independent parameters
(see ST03 for details).  As mentioned before, the displacement between
the simulated bulges and the K00b could be due to the fact that we 
are using a constant mass-to-light ratio for all bulges at all analyzed
redshifts. The fact that the simulated bulges are displaced toward
the elliptical regions could  indicate that, in these simulations,
we tend to reproduce mainly large bulges. A wider range of bulges is
expected to be given if Supernova energy feedback is corrected implemented.
However, there is currently 
no consistent Supernova feedback implementation in SPH (see
Scannapieco et al. 2006 and Tissera et al. 2006 for a promising
approach). 
In Table 1, we also included the corresponding linear fits for  disc structural parameters,
log $r_{\rm d}$ vs. $\Sigma^{\rm o}_{\rm d}$, which also show a good agreement with observations. 
 
\begin{table*}
\begin{minipage}[t]{\columnwidth}
  \caption{Correlations between mass structural parameters}
  \label{General}
\centering
\renewcommand{\footnoterule}{}  % to avoid a line before footnotes
\centering
\begin{tabular}{lcccccccccc}
\hline\hline
    & \multicolumn{2}{l}{$\Sigma_{\rm eff} $ vs. Log $r_{\rm eff} $} && 
\multicolumn{2}{c}{$\Sigma^{\rm o}_{\rm d}$ vs. Log $r_{\rm d} $}&&
\multicolumn{2}{c}{Log $n$ vs.$\Sigma^{\rm o}_{\rm b}$ }\\
%  \cline{2-3} \cline{5-6} \cline{8-9}
   & $a$ & $r$&& $a$ &  $r$ && $a$ &  $r$ &\footnote{Slopes of linear regressions ($y = ax + b$) and Pearson linear correlation coefficients
for the simulated galaxy-like objects at $z=0$ (Sim($z=0$)), for their progenitors during
merger events (Sim($z > 0$)), and the observed bulges and discs of Khosrhoshahi et al. (2000b).
%Standard dispersion are given within parenthesis.
}
\\
    \hline
Sim($z=0$)      &0.13 $\pm 0.01$ & 0.94 && 0.13 $\pm 0.02$ &  0.92 && $-0.09 \pm 0.02$ & $-0.77$ \\
Sim($ z >0$) 	&0.15 $\pm 0.01$ & 0.90  && 0.14 $\pm 0.01$ & 0.91 && $-0.10 \pm 0.01$ &  $-0.71$\\
K00b 		&0.15 $\pm 0.05$& 0.52 &&  0.24 $\pm 0.06$ & 0.65 && $-0.08 \pm 0.01$ & $-0.88$\\
%GG03 		&0.12 $\pm 0.03$ &0.72 && --- &---&&                $-0.06 \pm 0.01$  & $-0.80$ \\
\hline
  \end{tabular}
\end{minipage}
\end{table*}

\begin{table*}
\begin{minipage}[t]{\columnwidth}
  \caption{Mean variation in structural parameters and FP residuals}
  \label{bootstrap}
\centering
\renewcommand{\footnoterule}{}
\begin{tabular}{lcccccccccccc}
\hline\hline
&\multicolumn{2}{c}{$\Delta$FP}&\multicolumn{2}{c}{ $\Delta r_{\rm eff}$}& \multicolumn{2}{c}{$\Delta \Sigma_{eff}$}& 
\multicolumn{2}{c}{$\Delta \sigma_0$}& \multicolumn{2}{c}{$\Delta n$}& \multicolumn{2}{c}{$\Delta M_{\rm b}$\footnote{Statistical errors estimated by applying the bootstrap technique  are given within parenthesis.}}\\
\hline
&ODP & Fusion&ODP & Fusion&ODP & Fusion&ODP & Fusion&ODP & Fusion &ODP & Fusion\\
DSBs&-0.09 & 0.03& -0.26&-0.14&0.46&-1.02&25.6&25.1&-0.57&0.17&1.15&1.25  \\
Errors &0.03&0.01&0.12&0.10&0.53&0.33&6.97&6.10&0.22&0.21&0.42&0.44\\
SSBs&-0.01&0.03&0.05&0.03&0.28&0.02&-3.93&8.99&-0.12&0.01&-0.10&-0.11\\
Errors&0.01&0.05&0.03&0.04&0.19&0.83&0.90&7.72&0.10&0.47&0.23&0.56\\
\hline
\end{tabular}
\end{minipage}
\end{table*}

\subsection{The  Fundamental Mass Plane for simulated bulges}

As it was discussed in the Introduction, it is not yet clear if 
bulges determine the same FP of elliptical galaxies.
In this Sect., we investigate if the simulated bulges at $z=0$ satisfy a  FP mass  relation.

For that purpose, we  adopt  the luminosity FP 
determined by  FBPB02 for  Coma cluster ellipticals
and early-type spiral bulges as a reference plane:

\begin{equation}
FPR= 1.3 \cdot {\rm log}~\sigma_0 + 0.3 \cdot \Sigma_{\rm eff}-7.31,
\end{equation}

\noindent where $\sigma_0$ is in $\rm{km~s^{-1}}$ and $\Sigma_{\rm eff}$ is
in $\rm{mag~arcsec^{-2}}$.

We assume 
the circular velocity ($V^2 = GM/(r^2+\epsilon_g^2)^{3/2}$, where $\epsilon_g$ is the gravitation softening) at 2 kpc $h^{-1}$  as an estimate of
the depth of the potential well and central velocity dispersion ($\sigma_{0}$).
As we have already mentioned, the effective surface brightness ($\Sigma_{\rm eff}$) 
is obtained following McArthur et al. (2003) from $n$ and 
$\Sigma^{\rm o}_{\rm b}$.
%, the latter being calculated from the mass surface 
%density by assuming a constant mass-to-light ratio.
We acknowledge
that the mass-to-light ratio can change 
with luminosity and redshift, among other things.  
However, we keep it constant, so that
{\it deviations from the mass FP  can be directly linked
to  differences in  mass distributions and velocity patterns}, and not to 
stellar population evolution or dust effects.

\begin{figure}
\begin{center}
\vspace*{-0.2cm}\resizebox{7.5cm}{!}{\includegraphics{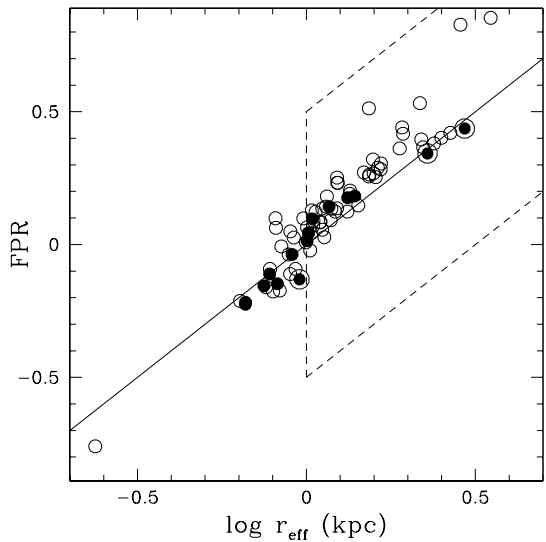}}\vspace*{-0.2cm}%
\end{center}
\caption{The simulated  fundamental mass plane for bulges
 at 
$z\approx 0$ (filled circles) and their progenitors during merger
events (open circles). For the sake of confrontation, constant mass-to-light ratios for the bulge and the disc components
have been assumed by requiring the $z=0$ systems to match observations.
Encircled filled circles correspond to the parameters for
the high resolution systems.
The observed  FP of FBPB02 
(solid line) and the region  occupied by ellipticals  (dashed rectangle) have
 been also  depicted.}
\label{PFfb02}
\end{figure}

In Fig.~\ref{PFfb02}, we show the FP of the simulated bulges   
at $z=0$ and their progenitors during merger events.
 We have also included the region
occupied by ellipticals taken from FBPB02. As expected, the simulated
bulges  populate the smaller $r_{\rm eff}$ extreme,
and at $z=0$, they show a slope and scatter in agreement with observations.
Their progenitor systems at $z >0 $ tend to be above the $z=0$ relation.

The HR systems follow the same  FP mass relation and require the same mass-to-light ratios to match the observations.
We claim that if systems at $z=0$ have structural parameters and reproduce a fundamental plane in 
agreement to those of the main runs, then their evolutionary history during mergers would be  
statistically similar. Hence,  for the sake of simplicity, we only include calculations for the
HR systems in the analysis at $z=0$.

\subsubsection{The effects of mergers}
\label{effects_of_mergers}

In this  section, we will focus on the analysis of the effects of
mergers on the  FP mass  relation by studying  mergers during two phases:
orbital decay  and fusion phases. In this way, we will try
to individualize the effects of secular evolution and collision. The merger events analyzed here are those studied
by ST03, as  was mentioned in Sect. 2.2.

\begin{figure}
\begin{center}
\vspace*{-0.2cm}\resizebox{7.5cm}{!}{\includegraphics{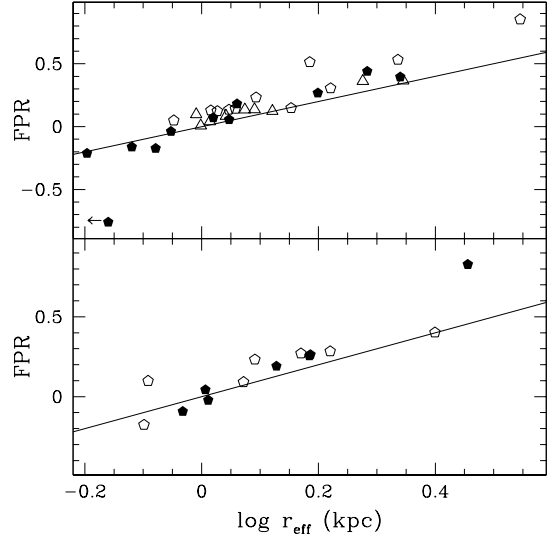}}\vspace*{-0.2cm}%
\end{center}
\caption{FP for simulated bulges 
undergoing double starburst  (upper panel) 
events at $z_{\rm A}$ (open pentagons), $z_{\rm C}$ 
(open triangles), and $z_{\rm D}$ (filled pentagons), and single 
starburst events (lower
panels)  at $z_{\rm A}$ (open pentagons)  and  $z_{\rm D}$ (filled pentagons).
The observed  FP of FBPB02
(solid line) has been included.  The pentagon with the arrow
depicts a point that, for the sake of clarity,  has been artifially displaced
from its original position at log $r_{\rm eff} =-0.62$ } 
\label{FPSD}
\end{figure}

In Fig.~\ref{FPSD}, we show the mass FP relations for 
systems at  double and single SB events.
The upper panel shows the mass FP defined at the three, defined stages of
the so-called double SB events. As it can be clearly seen, 
bulges at $z_{\rm A}$ (open pentagons), although being ordered,
 are all displaced to higher values than those of the observed FP at $z=0$ (solid line).
 After the secular evolution period (open triangles), bulges are
located more closely to the  observed relation.  Fusions 
(filled pentagons)  contribute to  scatter the objects
along this relation. 
When the systems experience single SB events (lower panel), their bulge 
components, on average,
already show  a  FP mass relation with a slope  in good agreement with observations at the beginning of the merger (open pentagons).
 The fusions do not seem to strongly modify the slope (filled pentagons).
We note, however, that one point is located  considerably far away 
from the relation after fusion. This system has not been detected as one that 
 experiences early gas inflows during the ODP,
 but it is gas-rich and transforms
an important fraction of its gas mass into stars during the fusion. Since
we are considering a  variety of merging systems from  cosmological
simulations, we expect not all the systems to follow exactly the same
patterns. 
%This particular system could be misclassified due to lack
%of adequate number of temporal outputs from the simulations.
An analysis of a larger set of mergers with higher numerical and temporal
resolution could show if the particular
 evolutionary path 
of this system is statitically significant. 

To quantify the changes in  FP relations during merger events,
 we followed FBPB02 and defined the residuals $\Delta FP$  along the x-axis by:

\begin{equation}
\Delta FP=({\rm log}~r_{\rm eff})_{\rm sim}-({\rm log}~r_{\rm eff})_{\rm obs},
\end{equation}

\noindent where $({\rm log}~r_{\rm eff})_{\rm sim}$ and $({\rm log}~r_{\rm eff})_{\rm obs}$ are the
effective radius estimated from the simulated bulge-disc decomposition and from the observed FP relation, respectively.

The assessment of  the effective changes in $\Delta FP$ during the two
main periods  in a 
merger event (i.e., ODP and fusion)
was done by  calculating the difference between the absolute values
 of the residuals during the orbital decay phase (i.e., $\Delta FP(z_{\rm C})$ and $\Delta FP(z_{\rm A})$)
and during the fusion itself (i.e.,
$\Delta FP(z_{\rm D})$ and $\Delta FP(z_{\rm C})$; for simplicity we assume $z_{\rm B}= z_{\rm C}$)
as a function of the variation in the structural parameters.
 We aim to detect if 
the net effect of the events is to strengthen the  FP relation or to 
dilute it by increasing the residuals.
Recall that  
double starbursts are  produced by secular evolution
induced by  the lack of well-formed bulges
that could provide stability to the discs (Tissera et al. 2002).
 Hence, this separation in SSBs and DSBs can
be interpreted as a classification of systems according to the
properties of their potential well.

As it can be appreciated from Table 2, 
 those systems which experience SSBs events do not significantly vary their 
structural parameters or their residual $\Delta FP$ during the ODPs.
Errors have been calculated by applying the bootstrap technique which provides a more reliable statistical
estimation of numerical noise due to low statistical number.
The variations in the residuals of the Fundamental Mass Plane are within a $\sigma$ level for these systems where a well-formed
bulge is present (i.e., $\Delta$FP $= -0.01 \pm 0.01$ during the ODP  and  $\Delta$FP $= 0.03 \pm 0.05$ during the fusion),
confirming that these objects are stable during interactions  and 
their mass distributions are not significantly altered 
during the secular evolution phase or in the fusion itself.
The structural parameters of these systems remain  statistically unchanged.

In those systems which undergo DSBs events, secular
evolution tends to reduce the residuals at a $3\sigma$ level ($\Delta$FP $= -0.09 \pm 0.03$),
while the fusion itself increases the dispersion also at a   $3\sigma$ level ($\Delta$FP $= -0.03 \pm 0.01$).
However, in absolute terms, secular evolution has a larger effect and contributes to tighten the relation.
From the analysis of the variation of the structural parameters of these systems, we found
that secular evolution helps to build up the bulge component and to make it more exponential
with a change in the mean $n$ shape parameter of $\Delta n = -0.57 \pm 0.22$ in agreement with previous results of
ST03. During both secular evolution and fusion, the bulges grow in mass and the potential wells
get more concentrated at a $3\sigma$level. Table 2 summarizes the variation in the parameters and residuals.
To summarize,  secular evolution seem to  contribute to form a bulge component and to diminish the 
residuals of the Fundamental Mass Plane relation as the bulge gets in place.

\begin{figure}
\begin{center}
\vspace*{-0.2cm}\resizebox{7.5cm}{!}{\includegraphics{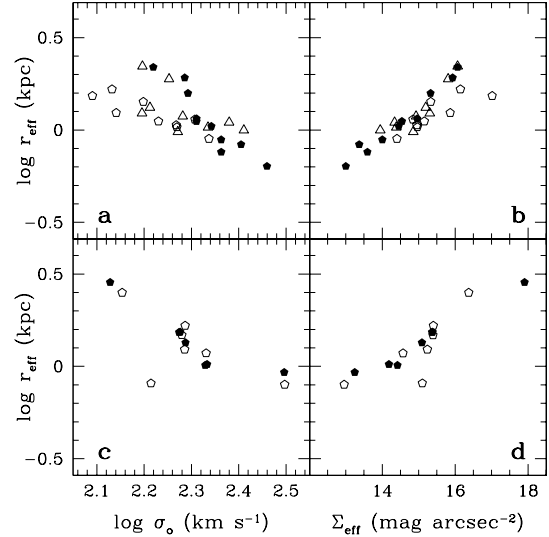}}\vspace*{-0.2cm}%
\end{center}
\caption{$Log(r_{\rm eff})$ as a function of $log(\sigma_{0})$ (a,c) and 
$\Sigma_{\rm eff}$ (b,d). Figures (a,b)  show the relations for the 
objects that experience double starburst events; meanwhile,
figures (c,d) 
correspond to the systems that present single starburst events. In all cases, 
open pentagons correspond to $z_{\rm A}$, open triangles to $z_{\rm C}$, and
filled pentagons to $z_{\rm D}$. }
\label{revssigmaSD}
\end{figure}

To improve our understanding of the changes occurring during 
secular evolution phases, we
plot the FP parameters against the structural ones at
the redshifts of interest.
Figure ~\ref{revssigmaSD}  
 shows log $r_{\rm eff}$ versus log $\sigma_{0}$ 
and $\Sigma_{\rm eff}$ (the so-called Kormendy relation).
In terms of the relation log$r_{\rm eff}$ versus 
log$\sigma_{0}$,
systems experiencing
DSBs finish the merger events with a much steeper slope: 
$d{\rm log}~r_{\rm eff}/d{\rm log}\sigma_{0}\approx -2.25 \pm 1.15$ ,
while they have  $ d{\rm log} r_{\rm eff}/d{\rm log}\sigma_{0} \approx -0.90 \pm 0.19$ at the beginning of 
the merger event (Fig.~\ref{revssigmaSD}a). 
In this case, fusions seem to contribute to the steepness of the slope so
that, at a given $r_{\rm eff}$, the simulated discs have larger
central velocity dispersion after the fusion. 
 In the case of SSB events, the systems show
an anti-correlation signal with $d{\rm log}~r_{\rm eff}/d{\rm log}\sigma_{0}\approx -1.38 \pm 0.30$,
   which is not affected by the fusion
(Fig.~\ref{revssigmaSD}c).

Finally, the simulated Kormendy relation for systems experiencing DSBs
adopts the observed slope (Table 1) after the secular evolution period,
which produces a change from $ d{\rm log} r_{\rm eff}/d\Sigma_{\rm eff} 
=  0.09 \pm 0.02$ at $z_{\rm A}$ to 
$  d{\rm log} r_{\rm eff}/d\Sigma_{\rm eff}= 0.16 \pm 0.03$ at $z_{\rm C}$ (Fig.~\ref{revssigmaSD}b).
In the case of bulges during SSBs, the correlations are in place with the observed slope when
the interactions start (Fig.~\ref{revssigmaSD}d) with $ d{\rm log} r_{\rm eff}/d\Sigma_{\rm eff}
= 0.14 \pm 0.04$. The ODP  or the fusion events
do not modify them.
%Systems that experience SSBs have 
% Kormendy's relation   consistent
%with the observed one before the interactions start.

Hence, in these simulations, it is   secular evolution  
 that  modifies the structural
parameters in such a way that their combination produces a  FP mass relation with the observed
slope.

\subsection{The  Tully-Fisher mass relation for GLOs}

Considering that the simulated bulges show structural parameters 
in agreement with those observed and that
mergers contribute to establish their fundamental relations, we would like to analyze whether the fundamental relation
for the disc components can be directly linked to the bulge ones.
For this purpose, we will study the TF mass relation  during merger events.
We have fitted a linear regression  to the 
 data of McArthur et al. (2003)  which comprise 304 nearby late-type (Sb-Sc) 
spiral galaxies in the
R band,

\begin{equation}
M_R=-6.36 \cdot [log~v_{\rm 2.2r_{\rm d}}-2.5]-21.5,
\end{equation}

\noindent where $v_{\rm 2.2r_{\rm d}}$ is  the velocity in ${\rm km~s^{-1}}$
measured at 2.2$r_{\rm d}$. 
The  magnitudes ($M_R$) of the GLOs are obtained 
assuming a constant mass-to-light ratio  of $\Gamma =7$ for the total baryonic mass, which
provides a consistent zero point for the $z=0$ relation. This ratio has been kept constant at all redshifts.

\begin{figure}
\begin{center}
\vspace*{-0.2cm}\resizebox{7.5cm}{!}{\includegraphics{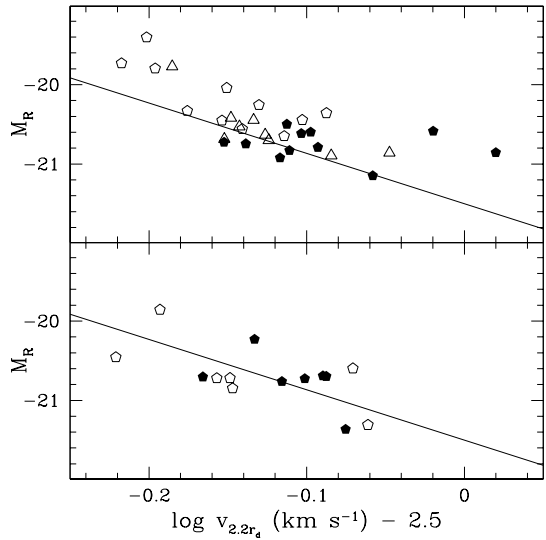}}\vspace*{-0.2cm}%
\end{center}
\caption{Tully-Fisher relation for simulated bulges 
undergoing double starburst (DS) events (upper panel) 
at $z_{\rm A}$ (open pentagons), $z_{\rm C}$ 
(open triangles), and $z_{\rm D}$ (filled pentagons) and single 
starburst (SS) events (lower
panels)  at $z_{\rm A}$ (open pentagons)  and  $z_{\rm D}$ (filled pentagons).
The solid line represents a linear regression fit to the observational
data of MacArthur et al. (2003).} 
\label{TFSD}
\end{figure}

Taking into account the results found for the bulges, we focus on the analysis of the effects of
fusions and secular evolution on the TF relation. In Fig.~\ref{TFSD}, we plot the TF relation for systems undergoing
DSBs (upper panel) and SSBs (lower panel)  at the redshifts of interest (i.e., $z_{\rm A}, z_{\rm C}, z_{\rm D}$).
Interestingly, we found that the TF relation behaves in a similar fashion to  the FP relation (see Fig.~\ref{FPSD}).
Those systems that experience secular evolution start systematically at a fainter magnitude than
that predicted by the observed local TF relation for a given velocity. After secular evolution, a larger fraction of
the simulated discs are consistent with the observed relation, are  more concentrated, and move along the
relation toward larger velocities and brighter magnitudes.
Fusions introduce scatter and help to displace the discs further in the same direction. 
In the case of systems undergoing SSB events, they all show a TF relation in agreement with observations.
Fusions tend to displace the systems along the TF relation toward higher velocities and brighter magnitudes, on average.
Again, we found that when a system is stable during an  orbital decay phase, it  already has
 structural parameters that
satisfy the fundamental relations for both the bulge and disc components.

The cross-correlation of  the parameters of the TF  and the FP relations
reveals that as discs grow in magnitudes (i.e., masses) and circular velocities, 
the bulges get more concentrated.  In particular, unstable systems
(i.e., those experiencing secular evolution) have
systematically less concentrated bulges and smaller discs.
Large discs grow as the bulge gets concentrated. 
%This behavior
%can be observed in Fig.~\ref{cross} where, as an example, we
%show the cross-correlation of the circular velocity of the disc components
%against the fundamental plane parameters during DSBs and SSBs.
Our results point  to  a coupled formation of both components, 
suggesting  
 a feedback process between them.

\section{Discussion and conclusions}

We have analyzed bulge-disc systems formed in hierarchical clustering
scenarios, finding that their structural parameters agree with those
of late-type bulges of spirals. 
We use the bulge-disc decomposition of the projected mass profiles to quantify {\it changes
in the mass distribution during mergers}.
The formation of systems with observational
counterparts is obtained by adopting an inefficient star formation which allows
the formation of compact bulges without exhausting the gas reservoir at early stages
of evolution. In this scenario, mergers do not completely destroy the discs, but
a remnant survives, leading to the formation of extended systems (Dom\'i{\i}nguez et al. 1998).
Recently, Springel \& Hernquist (2004) have shown that a correct modeling of the multiphase 
character of the interstellar medium produces similar results. Our simulations used a
phenomenological and simple scheme which  allows us to reproduce systems with  observational
counterparts and help to understand their evolution during mergers.

Note that our main aim is to study the changes during mergers and how GLOs get
to the relation at $z=0$ depending, on their evolutionary histories.
For the sake of simplicity, a constant mass-to-light ratio is assumed for all redshifts,
since we are interested in the analysis of the mass distributions.
 We have kept those values for all redshifts so that we can assess  the departures from 
the local relations  during violent events due to mass redistribution. 

We have analyzed a higher numerical resolution 
run  performed with Gadget-2, 
improving the resolution of baryons by one order of magnitude. We  found
similar trends in agreement with previous results from Dom\'i{\i}nguez et al. (1998).
This test proves that  the formation of galactic systems with the characteristics
described in this paper is independent of the code used and is not determined by 
numerical artifacts.

The following items summarize our results:

\begin{itemize}
\item The simulated bulges have a Kormendy mass relation with a slope
in agreement with observations at any redshift.
%, suggesting the possible existence
%of a unique Kormendy's relation for baryonic mass. 
However, when systems are segregated according to their internal properties, unstable systems (i.e.,
those that suffer double starbursts)
determine a shallower relation. The combination of systems at different stages of evolution 
 with different properties still
yields a relation with the correct slope, but with a larger dispersion. 

\item The simulated bulges show an anticorrelation
between the shape parameter $n$ and the central surface density 
$\Sigma^{\rm o}_{\rm b}$
in agreement with observations. Simulated bulges at $z=0$ are situated
at the faint end of the observed relation for intermediate and early-type spirals.

\item  At $z=0$, the simulated bulges define a Fundamental Mass Plane with a slope   consistent
with observations, assuming a constant mass-to-light ratio. Their progenitors
show a larger scatter. 
During gas-rich merger   events, unstable systems (i.e., those that undergo secular evolution during
their orbital decay phase) are located above the  Fundamental Mass Plane at $z=0$. After
the secular evolution phase, they are situated closer to the  Fundamental Mass Plane relation
at $3\ \sigma$ level. Fusions increase the scatter  along
the observed relation, but do not significantly change it.

\item We found that the parameters of the  TF mass relation are correlated
with those of the Fundamental Mass Plane in the sense that as the disc grows, the bulge
gets more concentrated.

\end{itemize}

By analyzing 
the changes in the mass distribution
of  galactic systems  during
merger events, we found
that secular evolution phases play an important role in the determination
of a  Fundamental Mass Plane   with a slope  in agreement with that of the  observed luminosity FP.
During these phases, gas inflows are driven,  triggering strong star
formation. In this way, compact stellar bulges are formed, and 
there is an increase in the  mass concentration and velocity dispersion.
We detect that the Kormendy relation for systems before experiencing
secular evolution is shallower than the observed one, and 
the slope of the FP relation tend to be steeper. After the secular phase, 
the simulated  Kormendy relation and the Fundamental Mass  Plane reproduce the observed
slope. The same behavior
can be observed between  $r_{\rm eff}$ and $\sigma_0$, so that
after secular evolution, the bulges  have higher central velocity
dispersion.
Those mergers that are not able to trigger gas inflows during the
orbital decay phase involve systems with well-formed bulges and   with structural
parameters that already determine relations with observational counterparts.
Taking into account previous works and, in particular, that of  Bender et al. (1992),
our findings suggest that secular evolution  seems to be
a necessary phase for the determination of a Fundamental Mass Plane with a slope in agreement with that  observed 
for  the luminosity FP.
In this sense, the different position  of compact 
and bright
dE on the FP could be in agreement  with the idea that 
the departure of  bright dwarf galaxies  might indicate 
a lack of an important secular evolution phase in their
history of formation.
Note that, in our simulations, systems get to the local fundamental relations
at different stages of evolution so that the  fundamental relations at
a given redshift are the result of the contributions of systems at different epochs of formation.
Hence, these differences in the merger histories introduce a natural dispersion in the relations
(see also Kobayashi 2005). 
Finally, we should note that we have mimiced the effects of SN feedback in the regulation of the star formation
activity by using a low star formation efficiency. Our results should be confirmed by models with self-consistent SN feedback treatment.

\begin{acknowledgements} 
High resolution simulations were run  on the  Ingeld PC cluster hosted by the Numerical Astrophysics group
at the Institute of Astronomy and Space Physics.
This work was partially supported by the
 Consejo Nacional de Investigaciones Cient\'{\i}ficas y T\'ecnicas,
  Fundaci\'on Antorchas,
and the European Union's ALFA-II
program, through LENAC, the Latin American European Network for
Astrophysics and Cosmology.

\end{acknowledgements}

%%%%%%%%%%%%%%%%%%

\clearpage

\end{document}